\documentclass[9pt,twocolumn,preprintnumbers,amsmath,amssymb,superscriptaddress,floatfix]{revtex4}

\usepackage{latexsym}
\usepackage{epsfig}
\usepackage{graphicx}
\usepackage{textcomp}
\usepackage{color}
\usepackage{amsmath}

\begin{document}

\title{Implementation of a Single-Shot Receiver for Quaternary Phase-Shift Keyed Coherent States}

\author{M. T. DiMario}
\affiliation{Center for Quantum Information and Control, University of New Mexico, Albuquerque, New Mexico 87131}
\author{E. Carrasco}
\affiliation{Center for Quantum Information and Control, University of New Mexico, Albuquerque, New Mexico 87131}
\author{R. A. Jackson}
\affiliation{Center for Quantum Information and Control, University of New Mexico, Albuquerque, New Mexico 87131}
\affiliation{Nanoscience and Microsystems Engineering, University of New Mexico, Albuquerque, New Mexico 87131}
\author{F. E. Becerra}
\email{fbecerra@unm.edu}
\affiliation{Center for Quantum Information and Control, University of New Mexico, Albuquerque, New Mexico 87131}


\begin{abstract}
{\color{black}We experimentally investigate a strategy to discriminate between quaternary phase-shift keyed coherent states based on single-shot measurements that is compatible with high-bandwidth communications. We extend previous theoretical work in single-shot measurements to include critical experimental parameters affecting the performance of practical implementations. Specifically, we investigate how the visibility of the optical displacement operations required in the strategy impact the achievable discrimination error probability, and identify the experimental requirements to outperform an ideal Heterodyne measurement. Our experimental implementation is optimized based on the experimental parameters and allows for the investigation of realistic single-shot measurements for multistate discrimination.}

\end{abstract}

\maketitle

\section{Introduction}

Discrimination of non-orthogonal states, such as coherent states, cannot be performed perfectly by any deterministic measurement \cite{helstrom76}. Measurements that minimize the probability of error are of fundamental interest, and have applications for information processing as well as quantum and classical communications \cite{bergou04, barnett09, gisin02, betti95, grosshans02}. The use of multiple coherent states to encode information can greatly increase the amount of information that can be transmitted in a single coherent state, and allow for communications with high spectral efficiency \cite{winzer12}. In addition, the non-orthogonality of coherent states makes them useful for secure communication via quantum key distribution \cite{bennett84, sych10}. However, this non-orthogonality is also detrimental to optical communication because it causes error in the measurements that decode the information \cite{helstrom76}, and limits the achievable rate of information transfer. Measurements for the discrimination of coherent states with sensitivities beyond the limits of conventional technologies have a large potential for optimizing some quantum communication protocols \cite{wittmann10} and enhancing optical communication rates \cite{chen12, becerra15}.

Discrimination measurements for two coherent states with different phases that can achieve sensitivities beyond conventional measurement limits have also been theoretically investigated \cite{taeoka08, dolinar73, kennedy72} and experimentally realized \cite{wittmann10, wittman08, cook07}. In a strategy proposed by Kennedy \cite{kennedy72} for the discrimination of two coherent states with opposite phases $\{|-\alpha \rangle, |\alpha\rangle \}$, the two input states are unconditionally displaced in phase space to the states $|0\rangle$ and $|2\alpha\rangle$, respectively, and photon counting is used to discriminate between these displaced states. This discrimination strategy achieves error probabilities below the optimal Gaussian measurement for two states, corresponding to the Homodyne measurement \cite{weedbrook12, taeoka08}, for a certain range of input powers. Furthermore, the amplitude of the displacement operation in phase space can be optimized to minimize the error probability \cite{taeoka08, wittman08}. This strategy, termed the optimized Kennedy strategy, achieves discrimination errors below the Homodyne Limit for all input powers.

Strategies for the discrimination of the four coherent states $|\alpha_{k}\rangle\in\{|\alpha\rangle,~|i\alpha\rangle,|-\alpha\rangle,|-i\alpha\rangle\}$  have been proposed \cite{bondurant93, izumi13, muller15, nair14, becerra11} and experimentally demonstrated \cite{becerra11, becerra13, becerra15, ferdinand17}. Based on displacement operations, photon counting, and fast feedback, these strategies achieve errors below the quantum noise limit (QNL) set by an ideal Heterodyne measurement. On the other hand, single-shot measurement strategies for multiple states based on displacement operations and single photon counting without feedback \cite{izumi12, kosloski12} are in principle compatible with high-bandwidth communication systems, and may provide advantages for high-speed quantum and classical communications.  One such strategy proposed by Izumi et. al. \cite{izumi12} uses single-shot measurements that perform multiple hypothesis testing simultaneously by multiple displacement operations. This strategy is a generalization of the optimized Kennedy measurement for four states, where the total error probability is minimized with optimized amplitudes of the displacement operations.

In this work, we experimentally demonstrate the strategy for the discrimination of four coherent states with single-shot measurements based on the work in Ref.\cite{izumi12}. The strategy can also be seen as a ``minimum resource" measurement as it does not require feedback or photon number resolution, and employs the minimum number of simultaneous measurements. We utilize an inherently phase-stable polarization-based setup to simultaneously test multiple hypotheses for the input state. Moreover, here we extend the previous theoretical work to include non-ideal visibility of the displacement operations, which quantifies imperfections encountered in any realistic implementation. The non-ideal visibility is one of the most critical parameters affecting the performance of the state discrimination process. Including this parameter, we observe an excellent agreement between our experimental observations and the theoretical predictions. This new model can be used as a guide for implementations of this single-shot multi-state discrimination strategy with current technologies. In Sec. 2 we describe the theoretical background for the strategy and our analysis including imperfect experimental parameters. In Sec. 3 we describe the implementation of the measurement and procedures, in Sec. 4 we present our experimental results, and we include our conclusions in Sec. 5.

\begin{figure}[!t]
\centering\includegraphics[width=\linewidth]{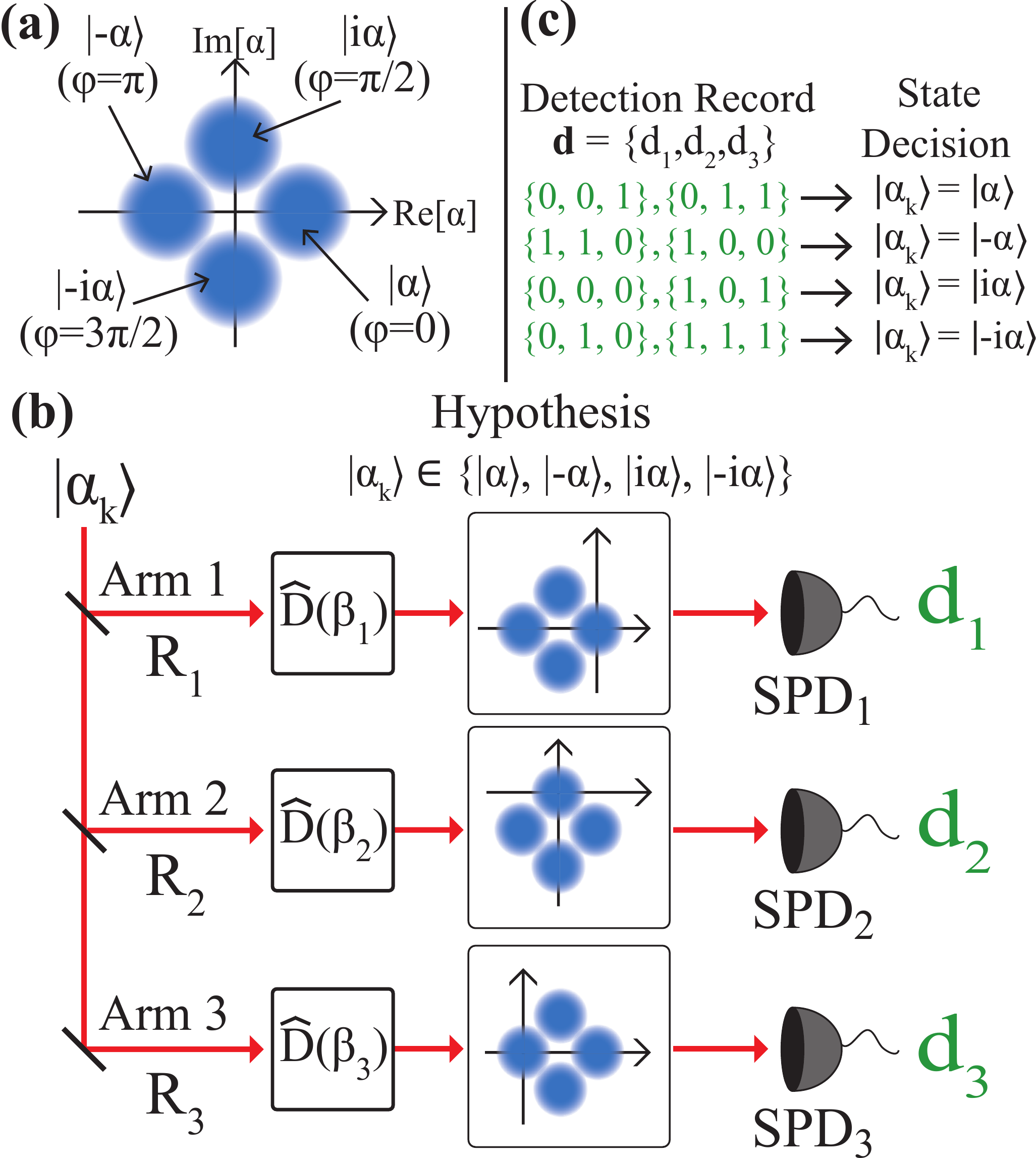}
\caption{ \label{A}\textbf{Theoretical Scheme:} (a) Input states are prepared in one of four possible states $|\alpha_{k}\rangle\in\{|\alpha\rangle,~|i\alpha\rangle,|-\alpha\rangle,|-i\alpha\rangle\}$.  (b) The power of the input state is split into three arms with ratios R$_{1}$, R$_{2}$, and R$_{3}$, and each arm is designed to test a different hypothesis of the input state by displacement operations and photon counting.  Arms 1, 2, and 3 are set to test states $|\alpha\rangle, |i\alpha\rangle, $ and $|-\alpha\rangle$, respectively, with optimized displacements $\hat{D}(\beta_{i})$ that minimize the probability of error.  The detection outcomes of three single photon detectors (SPD) for zero ($d_{i}=0$) or non-zero ($d_{i}=1$) photons provide information about the possible input state. These outcomes are used to obtain a decision rule based on maximum $a$ $posteriori$ probability criterion, which is shown in (c). }
\end{figure}

\section{Theoretical Background}

Figure 1 shows the schematic for the discrimination strategy of the four quadrature phase-shift-keyed (QPSK) coherent states with different phases $|\alpha_{k} \rangle = ||\alpha|e^{i\phi}\rangle$ (where $\phi = k\pi/2$ and $k \in \{0, 1, 2, 3\}$), as shown in Fig. 1(a), based on multiple hypothesis testing from Ref. \cite{izumi12}. The input state  $|\alpha_{k}\rangle$ is split into three separate detection arms as shown in Fig. 1(b), each of which tests a different hypothesis from the set $\{|\alpha\rangle,~|i\alpha\rangle,|-\alpha\rangle\}$ by displacing the input state and photon counting. The displacement operation $\hat{D}(\beta_{i})$ in arm ``$i$'' has a phase that is equal to the phase of the hypothesis being tested and magnitude set to an optimized value that minimizes the probability of error. The displacement of the input state in each arm $\hat{D}(\beta_{i})|\alpha_{k}\rangle$ is followed by photon counting with single photon detectors (SPD) with an on/off detection scheme \cite{izumi12}. The set of the three SPD outputs provide eight possible joint detection outcomes: $\textbf{d} = \{d_{1}, d_{2}, d_{3}\}  \in$ $\{$(1, 0, 0), (1, 0, 1), (1, 1, 0), (1, 1, 1), (0, 1, 1), (0, 1, 0), (0, 0, 1), (0, 0, 0)$\}$, where $d_{i}=0$ corresponds to detecting zero photons in arm ``$i$" and $d_{i}=1$ corresponds to detecting one or more photons. Based on the joint detection outcome $\textbf{d}$, the measurement uses the maximum $a$ $posteriori$ probability criterion as a decision rule (see Fig. 1(c)). The state with the highest joint posterior probability $P(\alpha_{k} | \textbf{d})$:

\begin{equation}
 P(\alpha_{k} | \textbf{d})=P(\alpha_{k}|d_{1},\beta_{1})P(\alpha_{k}|d_{2},\beta_{2})P(\alpha_{k}|d_{3},\beta_{3})
\end{equation}

is assumed to be the correct input state. Here $P(\alpha_{k} | \textbf{d})$ is the conditional posterior probability that the input state was $|\alpha_{k}\rangle$ given that the detection record $\textbf{d}$ was observed. $P(\alpha_{k}|d_{i},\beta_{i})$ is the conditional probability for state $|\alpha_{k}\rangle$ given the detection result of $d_{i}$ and a displacement amplitude $\beta_{i}$ in arm $``i"$, calculated through Bayes' theorem: $P(\alpha_{k}|d_{i},\beta_{i})P(d_{i}) = P(d_{i}|\alpha_{k},\beta_{i})P(\alpha_{k})$. Here $P(d_{i})$ is the probability of observing $\textbf{d}$ and $P(\alpha_{k})$ are the initial probabilities of the input states $\{|\alpha_{k} \rangle\}$, which are assumed to be equiprobable $P(\alpha_{k})=\frac{1}{4}$. The conditional probability of photon detection $P(d_{i}|\alpha_{k},\beta_{i})$ is given by the Poissonian probabilities:

\begin{equation}
P(d_{i}|\alpha_{k},\beta_{i})=
\begin{cases}
      e^{-|\alpha_{k}-\beta_{i}|^{2}}  , & d_{i}=0 \\
      1-e^{-|\alpha_{k}-\beta_{i}|^{2}}  , & d_{i}=1 \\
 \end{cases}
\end{equation}
\\
The total probability of error for this strategy is \cite{izumi12}:
\begin{equation}
P_{E} = 1 - \frac{1}{4}\sum_{k=1}^{4} P(\alpha_{k} | \alpha_{k}, \beta_{1}, \beta_{2}, \beta_{3})
\end{equation}
where $ P(\alpha_{k} | \alpha_{k}, \beta_{1}, \beta_{2}, \beta_{3})$ is the probability of correct discrimination given the input state $|\alpha_{k}\rangle$ was displaced by an amount $\beta_{1}, \beta_{2},$ and $\beta_{3}$ in arm 1, arm 2, and arm 3, respectively. Optimization of the displacement field amplitudes $|\beta_{i}|$ allows for the minimization of the probability of error in Eq. (3) given experimental imperfections and splitting ratios $R_{1}$, $R_{2}$, and $R_{3}$ (see Fig. 1b), which provide each arm ``$i$'' with an optimal fraction $R_{i}|\alpha|^{2}$ of the total input power $|\alpha|^{2}$ \cite{izumi12}.

Fig. 2(a) shows the numerically calculated optimal displacement ratios $|\beta_{opt, i}|^{2}/R_{i}|\alpha|^{2}$ for the three detection arms. Arm 2 (thin dashed blue line) has a higher optimal displacement ratio compared to arms 1 and 3 (solid red line). However, the optimal displacement ratios for all three arms converge to a 1:1 ratio as the input mean photon number $\langle n \rangle=|\alpha|^{2}$ increases. The inset (i) in Fig. 2(a) shows the optimal splitting ratio for each arm as a function of mean photon number. These ratios start at $\{R_{1}, R_{2}, R_{3}\}=\{0.33, 0.33, 0.33$\} but quickly converge to about $\{0.40, 0.20, 0.40$\}. We note that the state that arm 2 is testing for ($|\alpha_{k}\rangle=|i\alpha\rangle$) has the highest overlap with the other states being tested ($|\alpha\rangle, |-\alpha\rangle$). This  may be why the optimal splitting ratio results in a reduced amount of power that converges to 20\% of the total input power.

\begin{figure}[!htbp]
\centering\includegraphics[width=\linewidth]{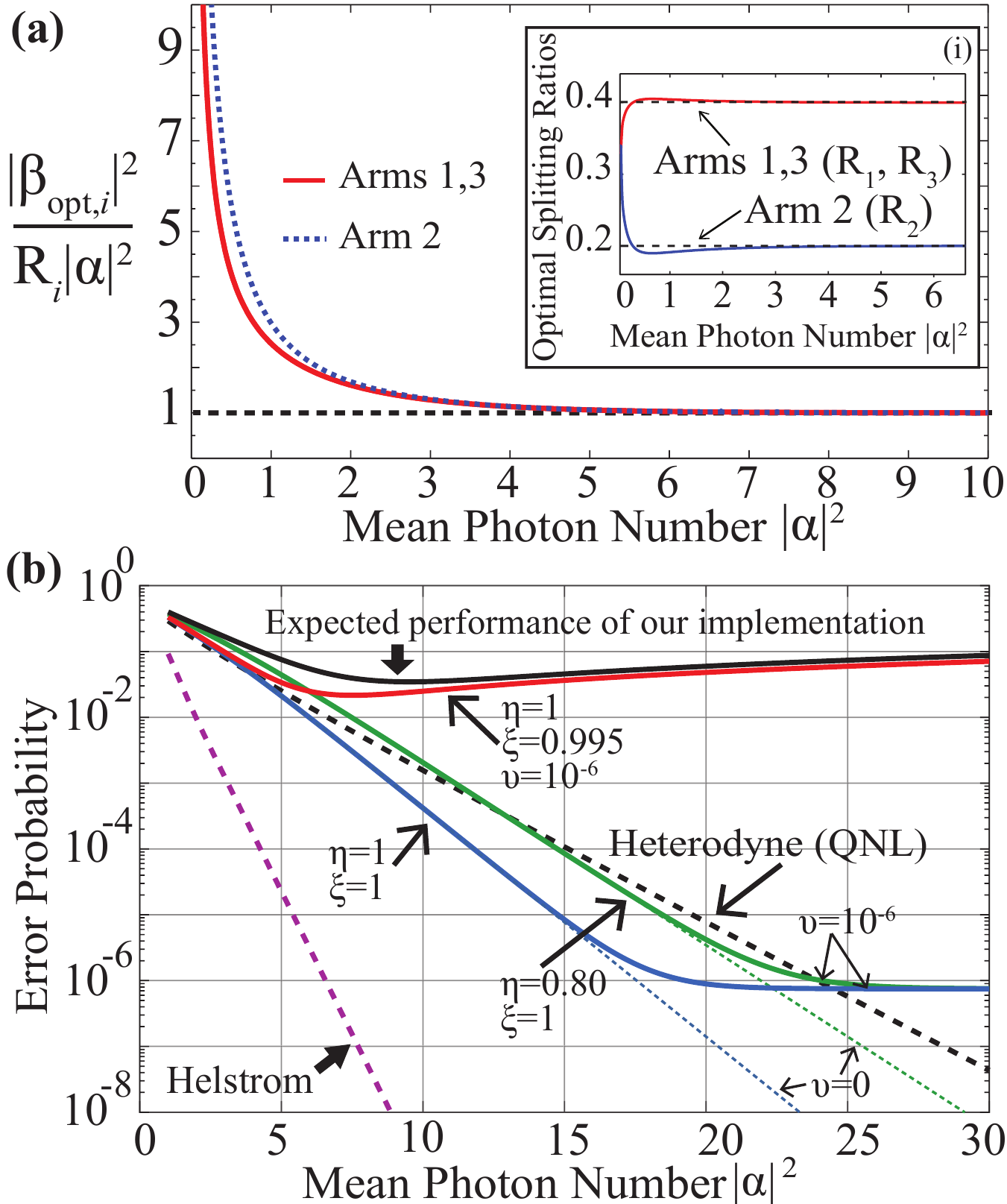}
\caption{\label{A} \textbf{Theoretical Optimal Parameters and Error Probabilities.} (a) Optimal displacement ratios ($|\beta_{opt, i}|^{2}/R_{i}|\alpha|^{2}$) for the three detection arms that minimize the probability of error. Arms 1 and 3 in a red solid line and arm 2 in a blue dashed line. The inset shows the optimal power ratios $R_{1}, R_{2}$, and $R_{3}$ in arms 1, 2 and 3, respectively. (b) Error probabilities for different values of detection efficiency ($\eta$), dark counts ($\nu$) and visibility ($\xi$) of the displacement operations for fixed splitting ratios of $\{R_{1}, R_{2}, R_{3}\} = \{0.40, 0.20, 0.40\}$ in arms 1, 2 and 3, respectively. Cases shown: detection efficiency of $\eta=1.0$, dark count rate of $\nu=10^{-6}$, and visibility of $\xi=$0.995 (red solid line); $\eta = 0.80$, $\nu = 10^{-6}$, with $\xi = 1.0$ (green solid line); and $\eta=1.0$, $\nu=10^{-6}$, with $\xi=1.0$ (blue solid line).  Thin dotted lines of the same color represent zero dark counts $\nu=0$.  The Helstrom bound, corresponding to the ultimate quantum limit \cite{helstrom76}, and Heterodyne Limit (QNL) are also plotted in thick dashed lines.}
\end{figure}

Following the theoretical work in Ref. \cite{izumi12}, we analyzed the discrimination strategy including realistic, non-ideal detection efficiency ($\eta$) and non-zero dark counts ($\nu$). Furthermore, we extend the work in \cite{izumi12} to incorporate imperfections in the displacement operations by allowing the interference of the two fields $(|\alpha_{k}\rangle, |\beta_{i}\rangle)$ to have non-ideal visibility ($\xi$), resulting from mode mismatch and other imperfections encountered in realistic implementations \cite{becerra11, becerra13, becerra15}. In our analysis we consider three detection arms, since no further improvements in the strategy are expected by increasing the number of arms \cite{kosloski12}. We only considered fixed values for the splitting ratios $R_{1}$, $R_{2}$, and $R_{3}$, which are set to the optimal splitting ratios of $\{0.40, 0.20, 0.40\}$ in arms 1, 2, and 3, respectively. These splitting ratios are near optimal for mean photon numbers larger than 3, which is the regime in which this single-shot strategy is expected to show advantages over the QNL.

Fig. 2(b) shows the probability of error as a function of mean photon number $|\alpha|^{2}$ for different values of the visibility of the displacement operations ($\xi$), SPD detection efficiency ($\eta$), and with or without dark counts ($\nu$) at a rate of $\nu=10^{-6}$ in each arm for fixed splitting ratios of $\{0.40, 0.20, 0.40\}$. As discussed in Ref. \cite{izumi12}, this single-shot strategy with an ideal detection efficiency and visibility outperforms a Heterodyne measurement at the QNL for $|\alpha|^{2} > 3$. Non-zero dark counts reduce the performance of this strategy at high $|\alpha|^{2}$ until it encounters a ``noise floor'' set by the dark count rate, which appears to be persistent for situations with reduced detection efficiency (see case with $\eta$=0.80 in Fig. 2b).

However, our analysis shows that the visibility of the displacement operations is a critical parameter, and even very small deviations from ideal visibility severely modify the performance of the strategy. We observe that even with perfect detection efficiency  $\eta$=1.0, a reduced visibility of $\xi$=0.995 prevents the probability of error from being below the QNL. Moreover, reduction in visibility imposes a stricter ``floor" on the probability of error than the one from non-zero dark counts. The error comes down to this floor and then increases slightly with $|\alpha|^{2}$. For reference at $|\alpha|^{2} = 15$, a visibility of $\xi = 0.99998$, and perfect detection efficiency, the floor of the error probability will still be slightly above the QNL. Other values of visibility and detection efficiency define regions in the parameter space where this strategy is expected to outperform the ideal Heterodyne measurement, as described in Sec. 4.

\section{Experiment}
\subsection{Experimental Setup}
Figure 3 shows the diagram of the experimental setup. Coherent states are prepared in 5.42 $\mu$s long pulses using a 633nm power-stabilized HeNe laser together with an acousto-optic modulator (AOM) at a rate of 11.7 kHz.  The light pulses are coupled into a single mode fiber (SMF) and then directed to the polarization-based interferometric set-up.  The horizontal polarization is defined as horizontal to the optical table. A half-wave plate (HWP$_{0}$) along with a quarter wave plate (QWP$_{0}$) are used to prepare the four possible input states $|\alpha_{k}\rangle \in \{|\alpha\rangle,|i\alpha\rangle,|-\alpha\rangle,|-i\alpha\rangle\}$ with phases $\phi \in \{0, \pi/2, \pi, 3\pi/2\}$, respectively in the vertical (V) polarization. The strong local oscillator field $|$LO$\rangle$ is defined in the horizontal polarization (H). HWP$_{0}$ is set at an angle relative to the horizontal so that 1\% of the total power is contained in the input state  $|\alpha_{k}\rangle$ in the vertical polarization (V) \cite{becerra13b} with a mean photon number $\langle n \rangle = |\alpha|^{2}$ that is calibrated with a transfer-standard calibrated detector \cite{gentile96}.

\begin{figure}[!t]
\centering\includegraphics[width=\linewidth]{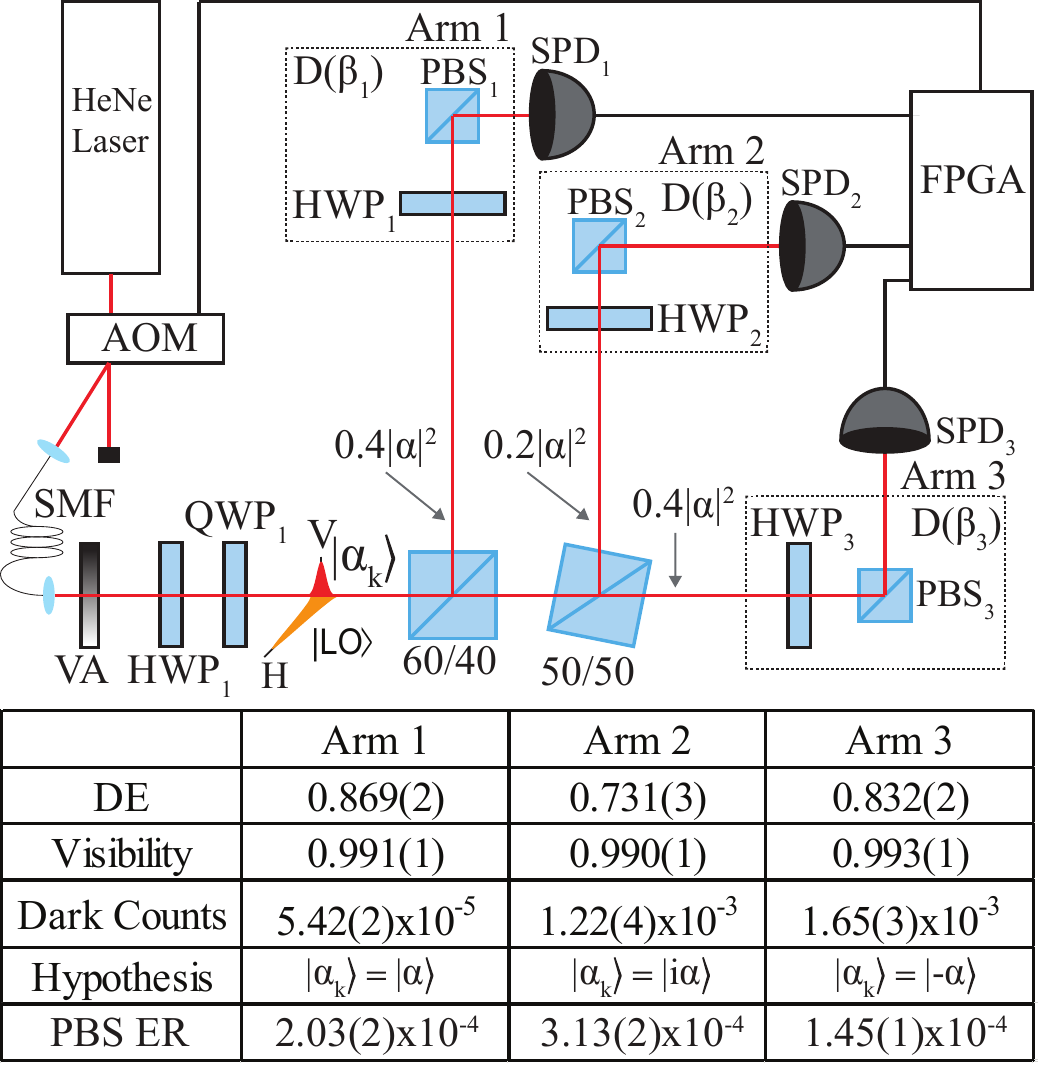}
\caption{\label{A} \textbf{Experimental Configuration.} The coherent state pulses are generated by a 633nm HeNe laser and an acousto-optic modulator (AOM) and then coupled into the experimental set-up using a single mode fiber (SMF).  A variable attenuator (VA) defines the mean photon number of the input state. A half-wave plate (HWP$_{0}$) and quarter-wave plate (QWP$_{0}$) define the four possible states in the vertical polarization which are then split into the three arms with a splitting ratio of $\{0.40, 0.20, 0.40\}$ with two beam splitters.  The input state in the vertical polarization co-propagates with the strong displacement field (LO) in the horizontal polarization. The displacements of the input state in each arm are performed by a HWP and a polarizing beam splitter (PBS) and the photons in the displaced state $\hat{D}(\beta_{i}) |\alpha_{k} \rangle$ are counted with SPD's.  The detection record $\bf{d}$ is collected by an field-programmable gate array (FPGA) which transfers the data to the computer for analysis of the discrimination strategy. The extinction ratio (ER) for each PBS$_{i}$ is shown and for a power ratio of the input state to LO of 1/100 this leads to the observed visibilities of $\{0.991(1), 0.990(1), 0.993(1)\}$ in arms 1, 2, and 3, respectively.}
\end{figure}

After the input state is prepared, the state is split into three detection arms with a 60/40 beam splitter and a 50/50 beam splitter. In order to obtain the desired splitting ratios of $\{0.40, 0.20, 0.40\}$ in arms 1, 2, and 3, respectively, the beam splitters are rotated to change the incident angles, which allows for some degree of tunability of the splitting ratios, such that the 50/50 beam splitter behaves as a 66/33 beam splitter. By using this technique, we achieve splitting ratios for the vertical polarization of:  $\{0.400(1), 0.200(1), 0.400(2)\}$.  We implement the optimized displacement operations $\hat{D}(\beta_{1}), \hat{D}(\beta_{2}),$ and $\hat{D}(\beta_{3})$ in each arm via polarization interferometry by using a HWP and a polarizing beam splitter (PBS) in each arm \cite{becerra13b}. Additional phase plates, not shown, are used in each arm to compensate for polarization rotation and birefringence. In arm ``$i$'', the half wave plate HWP$_{i}$ is set to an angle to produce interference between the input state $|\alpha_{k}\rangle$ and the displacement field $\beta_{i}(\langle n \rangle)$ on a polarizing beam splitter PBS$_{i}$ with a power ratio between these fields being $S_{i}(\langle n \rangle) =  |\beta_{i}(\langle n \rangle)|^{2}/R_{i}|\alpha|^{2}$, which correspond to the optimal displacements that minimize the probability of error $P_{E}$ in Eq. (3). Three SPDs detect the photons in the displaced states $\hat{D}(\beta_{i})|\alpha_{k} \rangle$ in each arm and a field-programmable gate array (FPGA) collects and transfers the joint detection result $\textbf{d}=\{d_{1}, d_{2}, d_{3}\}$ to a computer for analysis of the strategy as discussed in Sec. 2. The detection efficiency, visibility and dark counts for each arm are shown in Fig. 3. The extinction ratios for each PBS$_{i}$ are $\{2.03(2)$x$10^{-4}, 3.13(2)$x$10^{-4}, 1.45(1)$x$10^{-4}\},$ and for a power ratio of the input state to LO of 1/100 this leads to the observed visibilities of $\{0.991(1), 0.990(1), 0.993(1)\}$ in arms 1, 2, and 3, respectively.

\subsection{Calibration}
The implementation of the single-shot discrimination strategy requires precise calibration of the optimized displacement operations in the polarization basis. This calibration involves determination of the visibility of the displacement operation, and the angle of HWP$_{i}$ that provides the specific power ratio between the input field $R_{i}|\alpha|^{2}$ and the displacement field $|\beta_{i}|^{2}$ corresponding to the optimal displacements in arm ``$i$". As a first step, we determine the nulling angle $\psi_{i}$ of HWP$_{i}$ at which displacement to vacuum is achieved by total destructive interference at the output of the PBS$_{i}$ for $|\alpha \rangle$ in arm 1, $|i\alpha\rangle$ in arm 2, and $|-\alpha\rangle$ in arm 3. We then find the HWP angle that provides the level of interference corresponding to the optimal power ratio $S_{i}(\langle n \rangle)$ for each arm for different photon numbers. Appendix A describes the calibration procedure.

\begin{figure}[!t]
\centering\includegraphics[width=\linewidth]{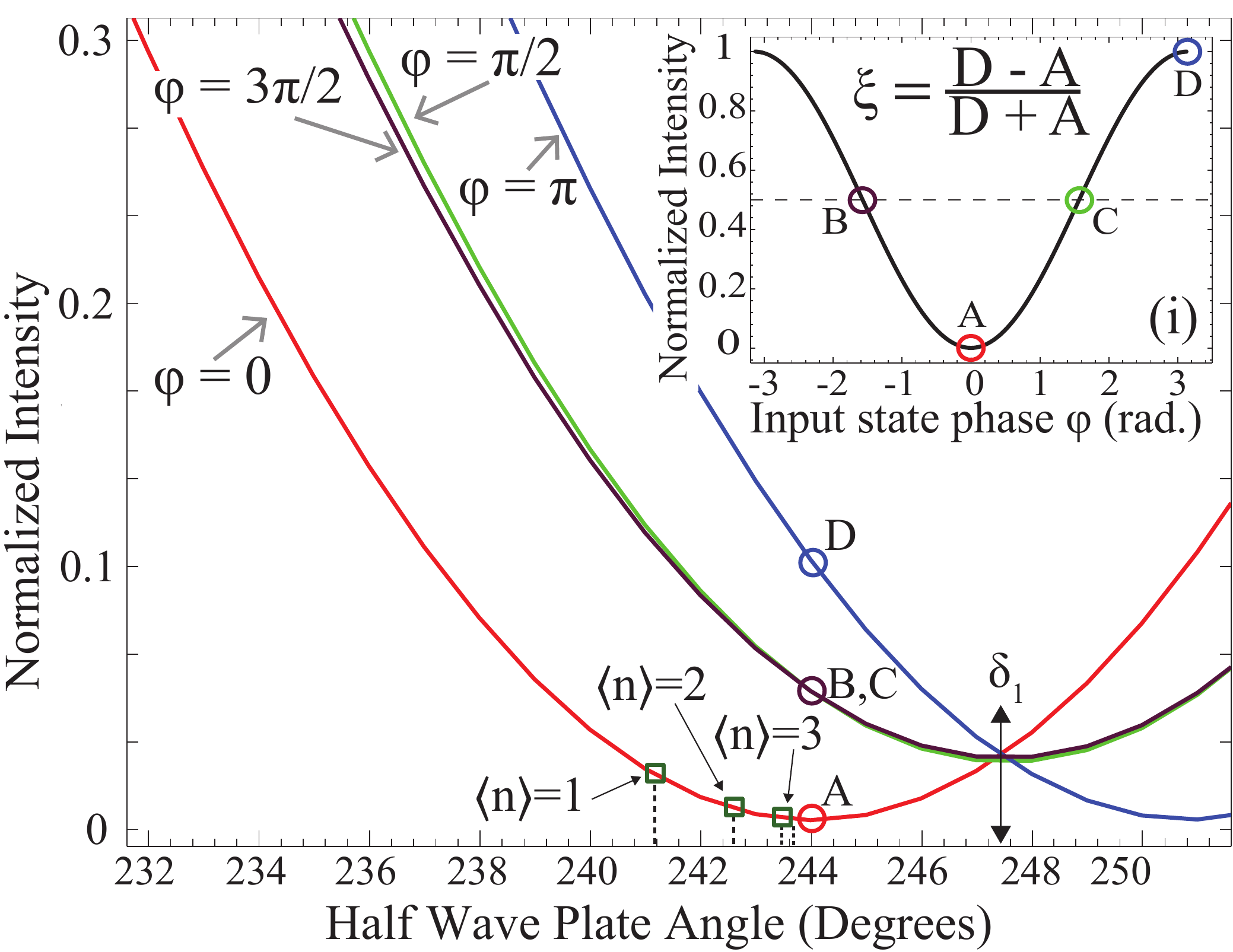}
\caption{\label{A} \textbf{Interference in Arm 1.} Normalized intensity after the PBS$_{1}$ as a function of HWP$_{1}$ angle in arm 1 testing the hypothesis that the input state is $|\alpha\rangle$ for the four possible input states $\{|\alpha\rangle,|i\alpha\rangle,|-\alpha\rangle,|-i\alpha\rangle\}$ with phases $\phi=\{0, \pi/2, \pi, 3\pi/2\}$, respectively.  Point A corresponds to the maximum nulling at the nulling angle $\psi_{1}$ and point D corresponds to the intensity at the nulling angle $\psi_{1}$ when sending the state with a $\pi$ phase shift ($\phi = \pi$).  These two points are the minimum (A) and the maximum (D) of an interference fringe and are used to estimate the visibility $\xi$ of the displacement operation in arm 1. The inset (i) shows the expected normalized interference as a function of input phase $\phi$ with minimum and maximum corresponding to points A and D, respectively. Points B and C correspond to HWP$_{1}$ being set to the nulling angle $\psi_{1}$ and sending the states $\phi =3\pi/2$ and $\phi =\pi/2$.  These two points are used to check the quality of state preparation. Also shown (green squares) are examples of the angles that correspond to the optimal displacement ratios  $S_{i}(\langle n \rangle)$ for $\langle n \rangle = 1, 2, 3$. The angle $\delta_{1}$ corresponds to the HWP$_{1}$ angle which allows only light with vertical polarization to be transmitted through the PBS$_{1}$.}
\end{figure}

Figure 4 shows an example for the calibration of the optimal displacements in arm 1, which is testing the hypothesis that the input state is $|\alpha\rangle$. The four curves show the normalized intensity after PBS$_{1}$ for a range of angles of HWP$_{1}$ for each input state with phases $\phi\in\{0, \pi/2, \pi, 3\pi/2\}$. Here, the red curve corresponds to the case when the input state is $|\alpha\rangle$ ($\phi=0$). A fit of the normalized intensity given by Eq. (A1) to this curve provides the determination of the nulling angle of $\psi_{1} \approx 244.0^{\circ}$ and the HWP offset of $\delta_{1}\approx247.2^{\circ}$. The angle $\delta_{1}$ corresponds to the HWP angle which allows only light with vertical polarization to be transmitted through the PBS$_{1}$. This information is used to calculate the angles $\theta_{1}(\langle n \rangle)$ of HWP$_{1}$ that give the optimal displacement ratio of $S_{1}(\langle n \rangle)$ for each mean photon number $\langle n \rangle$ (see Appendix A). The normalized intensities at the corresponding angles $\theta_{1}(\langle n \rangle)$ for the optimal displacement ratios for $\langle n \rangle={1, 2, 3}$ are shown as examples with green squares along the red curve ($\phi=0, |\alpha_{k}\rangle=|\alpha\rangle$).

Inset Fig. 4(i) shows the expected normalized interference in arm 1 after the PBS$_{1}$ as a function of the phase of the input state ``$\phi$" as a solid black line, and the measured intensities for the actual phases of the input states $\phi={0,\pi/2,\pi,3\pi/2}$ corresponding to the points (A, B, C, D) on the interference fringe, respectively, at the nulling angle for arm 1 ($\psi_{1}\approx244.0^{\circ}$). We estimate the visibility $\xi$ for each arm by using the intensity from the nulled state at point ``A" and the state that is shifted by $\pi$, shown by point ``D". We use the difference in intensity between sending the input state with phase $\phi=\pi$/2 , point ``C'', and the input state with $\phi=-\pi$/2, point ``B'', as an indicator to diagnose for imperfections and inaccuracies in the state preparation by HWP$_{0}$ and QWP$_{0}$. Our polarization-based experimental setup provides excellent phase stability and high accuracy for the implementation of the optimized displacements with low noise.

\begin{figure}[!b]
\centering\includegraphics[width=\linewidth]{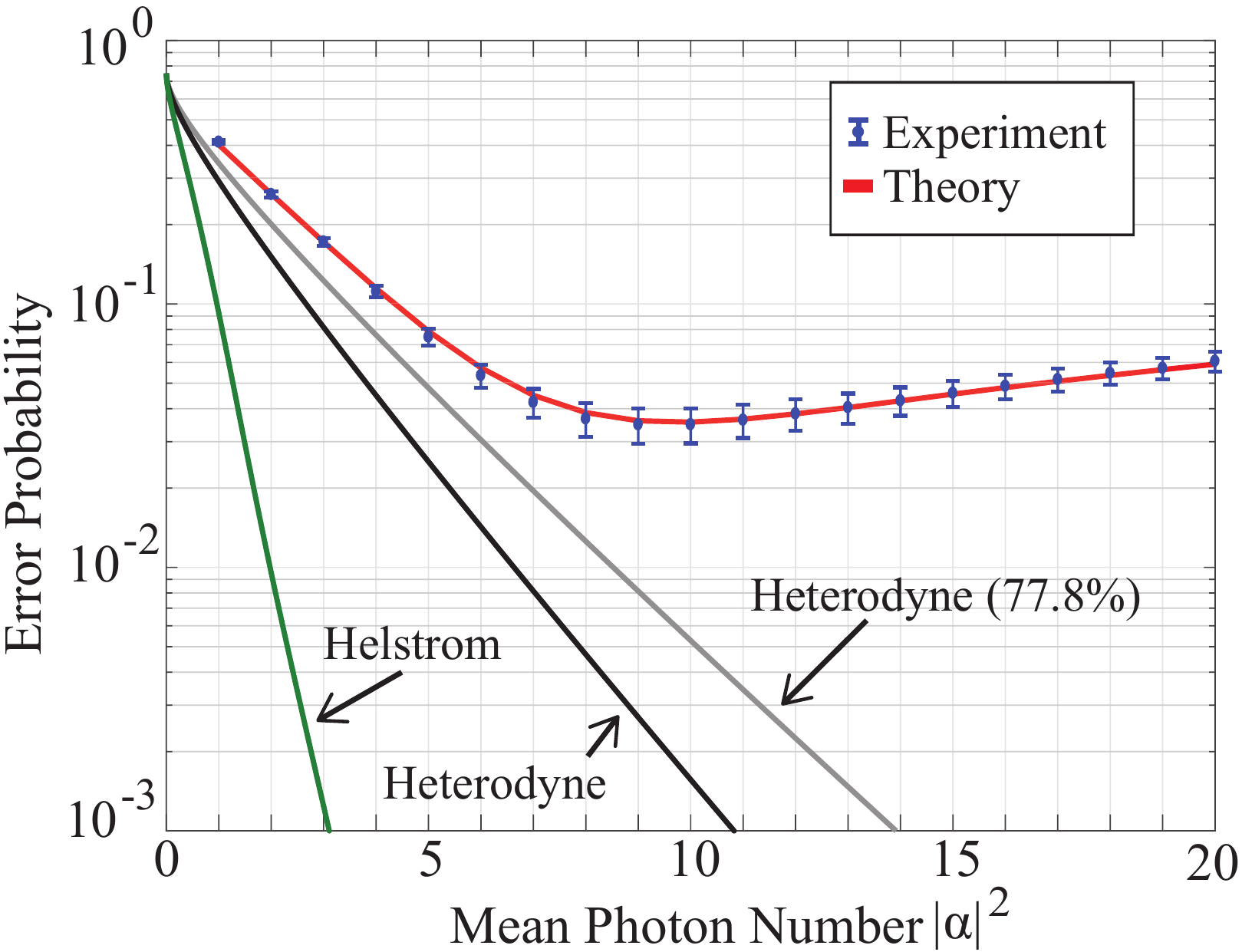}
\caption{ \textbf{Experimental Results.} Experimental results for the error probability as a function of mean photon number of the input state $\langle n \rangle = |\alpha|^{2}$. Included are the Helstrom Bound (Green) and Heterodyne Limit (QNL) (Black) as well as the QNL adjusted for the overall experimental detection efficiency of 77.8\% (Grey). The effect of non-ideal visibility can be seen by the floor that is imposed by the reduced visibility resulting in an achieved  $P_{E}$ of 3.6x10$^{-2}$ at a mean photon number of  $|\alpha|^{2}\approx10$. The experimental results are in very good agreement with theoretical predictions (Red). Our numerical study can provide a guide for future implementations of these single-shot strategies regarding the required experimental parameters needed to outperform a Heterodyne (QNL) measurement.}
\end{figure}

\begin{figure*}[!t]
\centering\includegraphics[width=.95\linewidth]{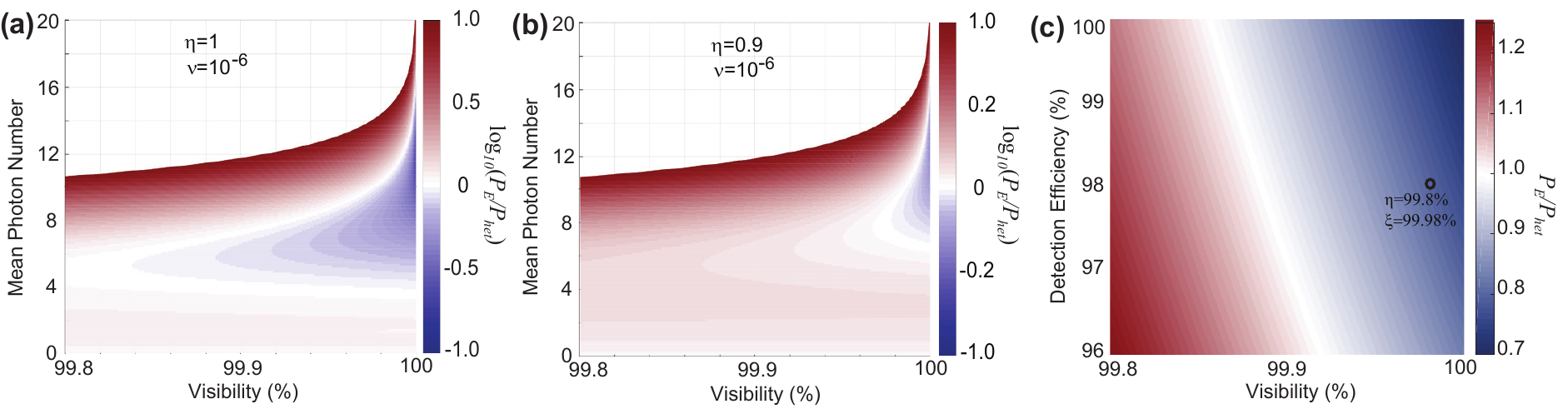}
\caption{ \textbf{Error probability for non-ideal parameters.}
Comparison of error probability $P_{E}$ for the non-Gaussian measurement with the Heterodyne limit $P_{het}$ in a logarithmic scale $\log_{10}(P_{E}/P_{het})$ as a function of visibility and mean photon number with (a) detection efficiency $\eta=1$ and (b) $\eta=90$ for $\nu=10^{-6}$. Negative values of $\log_{10}(P_{E}/P_{het})$ indicate improvement over Heterodyne, i.e. $P_{E}/P_{het}<1$. Empty areas above red boundaries correspond to values of $\log_{10}(P_{E}/P_{het})>1$. (c) Ratio $P_{E}/P_{het}$ in the parameter space for visibility and detection efficiency for $\langle n\rangle=6$. Note that it is possible to attain improvement over Heterodyne, $P_{E}/P_{het}<1$, with non-ideal detection efficiency and visibility. The black circle corresponds to a receiver with $\eta=98\%$ and $\xi=99.98\%$. }
\end{figure*}

\section{Results and Discussion}
Figure 5 shows the experimental results for the implementation of the single-shot discrimination measurement for the four non-orthogonal states $\{|\alpha\rangle , |i\alpha\rangle , |-\alpha\rangle, |-i\alpha\rangle\}$ with optimized displacements implemented with our polarization-based setup. Included in the plot are the theoretical predictions, the Helstrom bound \cite{helstrom76}, the Heterodyne Limit (QNL) for ideal detection efficiency, and the Heterodyne Limit with the same detection efficiency as our experimental implementation $\eta_{avg}=0.778(1)$, which includes optical losses in the set-up as well as detector efficiency in each arm.  We observe that our results lie slightly above the adjusted QNL (77.8\% detection efficiency) for low mean photon numbers and then encounter the floor which is set by the visibility in each arm as discussed in Sec. 2. Our experimental observations are in very good agreement with the theoretical predictions that include the critical parameters for state discrimination: detection efficiencies, SPD dark counts, and visibilities of the displacement operations. This indicates good fidelities in state preparation and implementations of optimized displacements in the separate detection arms.

This experimental setup allows for the robust implementation of the optimized displacements based on polarization-based interferometers, and complements previous experimental demonstrations with optimized displacements in the context of BPSK state discrimination \cite{wittmann10, wittman08}. Our experimental setup can be used for studies of more sophisticated strategies with photon number resolving detection \cite{kosloski12}, which may provide higher sensitivities and further robustness for realistic implementations, and could allow for performing polarization and intensity noise tracking \cite{bina17}. Furthermore, single-shot discrimination strategies of multiple states with polarization multiplexing can be used to improve the performance of different communication protocols based on multi-state measurements of coherent states \cite{croal16, arrazola15, arrazola16}, and could potentially be compatible with high-rate and high-spectral-efficiency communications.

Our results validate experimentally the single-shot multi-state discrimination strategy in Ref. \cite{izumi12} and provide an extension to incorporate the experimental imperfections resulting in the reduction of the visibility of the optimal displacements. As predicted and observed in the experimental results, the visibility of the displacement operation significantly affects the performance of the strategy. Moreover, our analysis allows us to determine what visibility is required to outperform a Heterodyne (QNL) measurement with detectors that have a reduced efficiency and dark counts.

Figs. 6(a) and 6(b) show our investigation of the expected ratio of the error probability $P_{E}$ for the non-Gaussian measurements to the Heterodyne measurement $P_{het}$ in logarithmic scale $\log_{10}(P_{E}/P_{het})$ as a function of visibility $\xi$ and mean photon number for different detection efficiencies $\eta$ with dark counts of $\nu=10^{-6}$. Blue regions for which $\log_{10}(P_{E}/P_{het})<0$ indicate improvement over Heterodyne with $P_{E}/P_{het}<1$, showing that non-Gaussian receivers with non-ideal visibility, detection efficiency, and dark counts can provide a real advantage over the ideal Heterodyne. Fig. 6(c) shows the ratio $P_{E}/P_{het}$ in the parameter space of visibility $\xi$ and detection efficiency $\eta$  for $\langle n\rangle=6$ and $\nu=10^{-6}$. We observe that there is a trade-off between $\xi$ and $\eta$ allowing different system configurations for receivers to reach the regime for which $P_{E}<P_{het}$. These studies allow for determining the requirements for implementations of receivers with non-ideal detectors and optical components to surpass the Heterodyne limit, and can be used to guide future demonstrations of these strategies.

As a concrete example, we note that current developments in superconducting detectors have demonstrated detection efficiencies $\eta\geq98\%$ and negligible dark counts $\nu\approx0$ \cite{fukuda11}. In addition, off-the-shelf high extinction ratio polarizers can reach extinction ratios of 10$^6$:1 \cite{bennett95} or higher \cite{moeller69}, resulting in displacement visibilities of $\xi\geq99.98\%$. The black circle in Fig. 6(c) corresponds to a non-Gaussian receiver based on these technologies with $\eta=98\%$ and $\xi=99.98\%$. This receiver has an expected advantage over the Heterodyne of about $20\%$, with $P_{E}/P_{het}=0.80$, at $\langle n\rangle=6$. This shows that it is possible to obtain advantages over the Heterodyne measurement based on single-shot measurements with current technologies.

\section{Conclusion}
We experimentally demonstrate a single-shot discrimination measurement for multiple non-orthogonal coherent states based on the work in Ref. \cite{izumi12}, which uses the minimum number of simultaneous measurements and the simplest photon detection method. Our setup based on polarization multiplexing provides high phase stability and readily allows for arbitrary displacements to be implemented with high accuracy using only wave-plates and beam splitters, and these new techniques can have applications in high-bandwidth communication protocols.

Our setup allows us to investigate the critical experimental parameters that affect the probability of error, such as the non-ideal visibility of the displacement operations and non-ideal detectors. Our numerical study can provide a guide for future implementations of these strategies to perform multi-state discrimination with single-shot measurements below the QNL in realistic scenarios. We expect that our work will motivate further research on investigating and demonstrating robust discrimination strategies compatible with high-bandwidth free space quantum communication with multiple states under realistic conditions of noise.

\section{Acknowledgements}

This work was supported by NSF Grant PHY-1653670 and PHY-1521016.

\section*{Appendix A: Calculation of HWP Angles for Optimal Displacements}
\setcounter{equation}{0}
\renewcommand{\theequation}{A{\arabic{equation}}}

The calibration of the HWP angles that perform the optimal displacements in each arm is achieved by first observing the interference after the PBS while scanning the HWP, and then finding the intensities that correspond to the optimal displacement ratios  $S_{i}(\langle n \rangle)$ = $|\beta_{opt, i}|/|\alpha_{i}|$ in arm $``i"$. Here $|\alpha_{ i}| = \sqrt{R_{i}}|\alpha|$cos$(2\Delta)$ is the signal amplitude after the PBS, $\Delta$ is the HWP angle, and $R_{i}|\alpha|^{2}$ is the signal power in that arm. The intensity after the PBS for arm $``i"$ as a function of HWP$_{i}$ rotation angle is given by:

\begin{eqnarray}
I_{i}(\alpha, \beta, \phi, \Delta) = R_{i}|\alpha|^{2} \text{cos}^{2}(2\Delta) + |\beta_{i}|^{2}\text{sin}^{2}(2\Delta)\\
\nonumber
- 2\sqrt{R_{i}}|\alpha||\beta_{i}|\text{sin}(2\Delta)\text{cos}(2\Delta)\text{cos}(\gamma)
\end{eqnarray}

Here $\gamma$ is the relative phase between the input signal $|\alpha_{k}\rangle$ and the local oscillator $|\beta_{i}\rangle$ in arm $``i$'', and $\Delta = \theta-\delta_{i}$ corresponds to the HWP rotation angle $\theta$ being shifted by an offset $\delta_{i}$. This offset $\delta_{i}$ corresponds to the angle of the HWP$_{i}$ that rotates the polarization such that only the signal in the V polarization is transmitted through the PBS$_{i}$, which can be seen in Fig. (4) as the point where the intensity is the same for all input states. The intensity given by Eq. (A1) can be normalized to the power of the input state $R_{i}|\alpha|^{2}$ in the V polarization before the PBS$_{i}$ in arm $``i"$ such that $I^{norm}_{i}=I_{i}/R_{i}|\alpha|^{2}$:

\begin{eqnarray}
I_{i}^{norm}(\alpha, \beta, \phi, \Delta) = \text{cos}(2\Delta)^{2} + f_{i}^{2}\text{sin}(2\Delta)^{2} \\
\nonumber
- 2f_{i}\text{sin}(2\Delta)\text{cos}(2\Delta)\text{cos}(\gamma)
\end{eqnarray}
\

where $|\beta_{i}|^{2}$ is the power of the LO in the H polarization before the PBS in arm $``i$" and $f_{i}$ is the ratio of LO amplitude to input amplitude $f_{i}=|\beta_{i} |/\sqrt{R_{i}}|\alpha|$ in arm $``i"$ before the PBS. Given the parameters $f_{i}$ and $\delta_{i}$, the expected intensity $I_{i}$ after PBS$_{i}$ in arm $``i"$, and the target optimal ratio $S_{i}(\langle n \rangle)$ = $|\beta_{opt, i}|/|\alpha_{ i}|$ of the LO  to the input field after the PBS for a given $\langle n \rangle$, we find the HWP angle $\theta_{i}(\langle n \rangle)$ that implements the optimal displacement in arm $``i"$ as:

\begin{align}
S_{i}(\langle n \rangle) &= \frac{|\beta_{opt,i}(\theta, \delta_{i})|}{|\alpha^{'}_{ i}(\theta, \delta_{i})|} \\
\nonumber
\\
\nonumber
&= \frac{|\beta_{i}|\text{sin}[2(\theta_{i}(\langle n \rangle) - \delta_{i})]}{\sqrt{R_{i}}|\alpha|\text{cos}[2(\theta_{i}(\langle n \rangle) - \delta_{i})]}   \\
\nonumber
\\
\nonumber
 &=f_{i}\text{tan}[2(\theta_{i}(\langle n \rangle) - \delta_{i})]
\end{align}

This procedure results is a list of angles $\theta_{i}(\langle n \rangle)$ for the HWP in each arm corresponding to the optimal ratio $S_{i}(\langle n \rangle)$ of the two fields for different $\langle n \rangle$.

\end{document}